\documentclass[twocolumn,showpacs]{revtex4}
\usepackage{graphicx}
\usepackage{bm}% bold math
\begin{document}

\title{Comment on ``Time-averaged properties of unstable periodic
orbits and chaotic orbits in ordinary differential equation systems''}
\author{Michael A. Zaks\email{zaks@physik.hu-berlin.de}}
\affiliation{ Institut f\"ur Physik, Humboldt-Universit\"at zu Berlin,
D-12489 Berlin, Germany }
\author{Denis S. Goldobin \email{denis.goldobin@gmail.com}}
\affiliation
{Department of Theoretical Physics, Perm State University, 614990 Perm, Russia}

\date{\today}
\begin{abstract}
The recent paper claims that mean characteristics
of chaotic orbits differ from the corresponding values averaged
over the set of unstable periodic orbits, embedded 
in the chaotic attractor. 
We demonstrate that the alleged discrepancy is an artifact of
the improper averaging: Since the natural measure is
non-uniformly distributed over the attractor, different periodic orbits
make different contributions into the time averages. 
As soon as the corresponding weights are
accounted for, the discrepancy disappears.
 
\end{abstract}

\maketitle

Recent Rapid Communication \cite{Saiki_Yamada} compares
properties of periodic orbits (UPOs) embedded
into a chaotic attractor to those of chaotic trajectories on this
attractor. Analysis is based on numerical data, and culminates in
the statement: ``time-averaged properties along a set of UPOs and
a set of chaotic orbits with finite lengths are totally different
from each other''. It is further conjectured that ``the time averages
of the dynamical quantities along UPOs with the same period of the 
Poincar\'e map have a limiting distribution with nonzero variance''.
In this comment we show that under the proper averaging procedure,
there seems to be neither the ``total'' difference between the averages, 
nor an argument for the above conjecture.  

Due to ergodicity, the value of time average for an observable
$A$ converges to $\overline{A}=\int A({\bm x})\mu({\bm x}) d{\bm x}$
where $\mu$ is the natural measure, and
integration is performed over the whole attractor. 
Approximation of  $\overline{A}$
by summation over the set of UPOs -- apparently the
method employed in~\cite{Saiki_Yamada} --
provides correct results only in exceptional cases
when $\mu$ is uniformly distrubuted over the attractor: 
for linear mappings like the Bernoulli map, symmetric tent map etc.
In general, however, the density of the natural measure
varies along the attractor. As a consequence, a chaotic orbit 
does not walk uniformly over the attracting set: it visits certain regions
relatively often or stays there relatively long. Contribution of such regions 
into the time averages is larger than of those visited seldom.  
As shown in \cite{Grebogi_Ott_Yorke_88} 
(see also Chapter 9.5 of the textbook \cite{Ott_book}),
non-uniform distribution of the natural measure can be recovered
from the properties of UPOs embedded in the attractor. In particular,
for invertible two-dimensional maps (and hence for three-dimensional
flows which induce such maps), the weight with which an UPO
contributes to the time average, is inversely proportional to 
the largest eigenvalue of the corresponding fixed point. Below
we demonstrate that taken the weights into account, 
the discrepancy between the mean values from the chaotic timeseries
and the mean values from the UPOs disappears.

We restrict ourselves to the first example considered in~\cite{Saiki_Yamada}: 
attractor in the celebrated Lorenz equations
\begin{equation}
\frac{dx}{dt}=\sigma(y-x),\;\;\frac{dy}{dt}=r x-y-x z,\;\;\frac{dz}{dt}=x y-b z
\end{equation}
with $\sigma=10,\;r=28, b=8/3$~\cite{Lorenz63}, and take
the same observable: the value of the variable~$z$.
In fact, already in \cite{Eckhardt-Ott-94} fractal
characteristics of the Lorenz attractor were evaluated
with the help of the properly weighted UPO data.
Our results are obtained by averaging $z$ for each orbit 
from the complete set of 111014 UPOs with $N\leq 20$ turns in the phase space 
around either of the attractor ``wings''. 
For each $N\leq 20$ we also computed $10^6$ segments of
chaotic trajectories with $N$ turns, and calculated values of 
$\langle z\rangle$ for every such segment.  

Comparison of eigenvalues for different UPOs with the same $N$ 
discloses strong inhomogeneity in the distribution of 
natural measure. Already among 186 orbits with $N=11$ the largest (4618.57)
and the smallest (415.59) eigenvalues differ by the factor of 11.1. 
At $N=19$ there are 27594 UPOs, and the ratio between the 
extremal eigenvalues is 77.2. 
Accordingly, the contribution of the ``most unstable'' UPO is hardly 
discernible, compared to the contribution of the ``least unstable'' one. 

\begin{figure}
%\centerline{\includegraphics[width=0.96\textwidth]{fig_1.eps}}
\centerline{\includegraphics[width=0.5\textwidth]{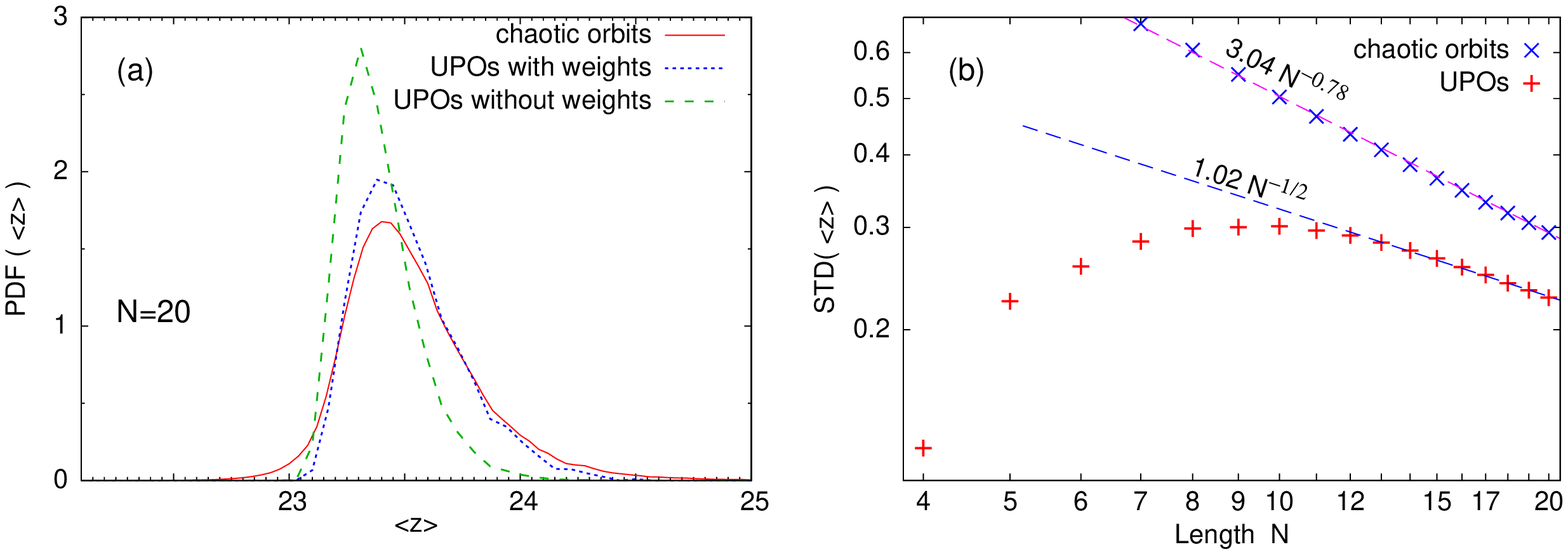}}
\caption{Time averages of $z(t)$ for chaotic trajectories and
UPOs of the Lorenz equations.\protect\\
(a) Probability density for orbits of length $N$=20.
Solid line: average values from $10^6$ segments of chaotic
trajectories. Dotted line: weighted means over all UPOs
with $N=20$. Dashed line: summation (without weights)
over all UPOs with $N=20$.\protect\\
(b) Standard deviation of $\langle z\rangle$ for 
periodic (pluses) and chaotic (crosses) orbits with $N$ turns.
Dashed lines: power-law fits.}
\label{fig_1}
\end{figure}

As seen in the left panel of Fig.\ref{fig_1}, taking the weights into account
shifts and re-shapes the distribution of mean values. The solid curve shows
the bell-shaped distribution of $\langle z\rangle$ for segments of chaotic orbits with length
20. The dashed curve shows the histogram obtained by summation of $\langle z\rangle$
from all UPOs of the same length. Similarly to Fig.1 of ~\cite{Saiki_Yamada},
the maxima of these two curves are shifted with respect to each other. 
The expectation values of $\langle z\rangle$ for these two distributions are distinctly different: 
$\overline{\langle z\rangle}$ equals 23.555 for chaotic trajectories and 23.420 for  
summation over UPOs. This difference, however, almost vanishes for the histogram 
which incorporates the weights of UPOs: the dotted curve
in Fig.\ref{fig_1}(a) is much closer to the solid curve, and has
$\overline{\langle z\rangle}$=23.554.  In the case
of shorter orbits with length $N=11$ reported in~\cite{Saiki_Yamada}
the same effect takes place:
ensemble of chaotic orbits predicts $\overline{\langle z\rangle}$=23.562, 
which is definitely distinct
from the value 23.420 obtained by summation over UPOs, but is much
closer to the value 23.550 yielded by proper summation with weights. 

Obviously, the distribution obtained from finite segments of chaotic 
orbits cannot be identical to the distribution produced by the set of UPOs 
of the same length. $N$ rotations in a piece of a chaotic orbit 
do not necessarily constitute a visit into the neighborhood of an UPO 
of the same length: they can consist of several visits into vicinities 
of the shorter UPOs, be a part of the passage near the longer UPO, etc.
Therefore -- at least in the checked range of $N$ -- distributions based on
chaotic orbits are broader than their counterparts built from the UPOs data.
As shown in the right panel of Fig.\ref{fig_1}, the breadth of both distributions
decays as a power of $1/N$. 
For chaotic segments  it approaches the law $3.04\,N^{-0.78}$
(which nearly coincides with the values from~\cite{Saiki_Yamada}), whereas the
standard deviation for a distribution from the set of UPOs with properly
assigned weights decays as $~1/\sqrt{N}$. Of course, the employed values of $N$
are, at best, moderate, and one cannot judge on the asymptotical properties of
the dependence. At any rate, however, these data unambiguously show
that in the range $N\leq 20$ there are no arguments against eventual convergence
to the $\delta$-shaped distribution\footnote{In fact, 
even for the set of unweighted UPOs, the standard deviation 
of $\langle z \rangle$  decays as $\sim 0.74/\sqrt{N}$. This also
implies convergence to the $\delta$-shaped distribution, -- centered, 
of course, {\em not} at the value of the time average.}.

Since for small and moderate values of $N$ the distributions are typically broad, 
it hardly makes sense to discuss whether an attractor of a particular set
of equations is accurately modeled by a single UPO of the given length: some of the
UPOs are definitely inappropriate, whereas some others may prove to be good.   
Average values computed along the UPO match the averages along chaotic
orbits either in a pathological case when all UPOs except one possess 
giant eigenvalues (and, hence, negligible statistical weights), 
or if one deliberately chooses the UPO whose
characteristics are close to that of the whole ensemble. For the latter,
however, the ensemble (or, at least, its representative parts) should
be examined, so there is hardly a gain in the computational efficiency.   
  
The last remark concerns estimates of the topological entropy
in~\cite{Saiki_Yamada}. It is known (see e.g. \cite{Sparrow82}) 
that at $r=28$ a chaotic orbit can make not more than 25 consecutive turns
around one wing of the Lorenz attractor before a jump to another wing.
Once each turn in the half-space $x>0$  is coded by "1" 
and each code in the complementary half-space $x<0$  is coded by "0", 
all binary strings with length $N\leq 25$ are met in the code
of a sufficiently long chaotic trajectory. 
The only missing periodic orbits are those whose
symbolic labels consist exclusively of ones or of zeroes. 
Accordingly, the number of UPOs with the length $N$ for $N\leq 25$
is given by the recursive formula $K(N)=\left(2^N-2-\sum_j j\,K(j)\right)/N$,
summation being taken along all divisors $j$ of $N$. 
However, estimate of the topological entropy as 
$h_{top}=\lim_{N\to\infty}\sup N^{-1}\log K(N)=\log 2$ is applicable 
only for the newborn attractor at $r=24.06\ldots$ in which
all symbolic strings of arbitrarily large $N$ are encountered. 
This is not the case for $r$=28: upwards from $N$=25 
the tree of symbolic sequences is ``pruned'', 
and the growth of number of UPOs as a function of $N$ may become slower. 
In any case, the range of orbit length $N\leq 14$ employed 
in~\cite{Saiki_Yamada} for the evaluation of the $h_{top}$ is far too short
and hardly appropriate for reliable estimates.

\end{document}